\documentclass[twoside]{article}
\usepackage[dvips]{graphicx}
\usepackage{qic,epsfig}
\usepackage{mathdots}
\usepackage{color}
\usepackage{ulem}
\usepackage{bm}
\usepackage{amsmath,amssymb}
\usepackage{type1cm}

\newtheorem{proposition}{Proposition}

\renewcommand{\thefootnote}{\fnsymbol{footnote}}  %use symbolic footnote
\DeclareSymbolFont{lettersA}{U}{txmia}{m}{it}
 \DeclareMathSymbol{\Indi}{\mathord}{lettersA}{'211}
 \usepackage{latexsym}%%%%%%% starting the text file 
\begin{document}
\setlength{\textheight}{8.0truein}    %FOR 2ND PAGE ONWARDS

\runninghead{Stationary measure for three-state quantum walk}
            {Takako Endo, Takashi Komatsu, Norio Konno, Tomoyuki Terada}
          
\normalsize\textlineskip
\thispagestyle{empty}
\setcounter{page}{1}

%\copyrightheading{Vol.}{No.}{Year}{Page Nos.}

\vspace*{0.88truein}

\alphfootnote

\fpage{1}
\centerline{\bf Stationary measure for three-state quantum walks}
\vspace*{0.37truein}
\centerline{\footnotesize
TAKAKO ENDO\footnote{endo-takako-sr@ynu.ac.jp}}
\vspace*{0.015truein}
\centerline{\footnotesize\it Department of Applied Mathematics, Faculty of Engineering,}
\baselineskip=10pt
\centerline{\footnotesize\it Yokohama National University, Hodogaya, Yokohama 240-8501, Japan}
\vspace*{10pt}
\centerline{\footnotesize
TAKASHI KOMATSU\footnote{
pt121199vy@kanagawa-u.ac.jp}}
\vspace*{0.015truein}
\centerline{\footnotesize\it Department of Mathematics and Physics, Faculty of Science, }
\baselineskip=10pt
\centerline{\footnotesize\it Kanagawa University
2946 Tsuchiya, Hiratsukashi, Kanagawa, 259-1293, Japan}
\vspace*{10pt}
\centerline{\footnotesize
NORIO KONNO\footnote{konno@ynu.ac.jp}}
\vspace*{0.015truein}
\centerline{\footnotesize\it Department of Applied Mathematics, Faculty of Engineering,}
\baselineskip=10pt
\centerline{\footnotesize\it Yokohama National University, Hodogaya, Yokohama 240-8501, Japan}
\vspace*{10pt}
\centerline{\footnotesize
TOMOYUKI TERADA\footnote{t.terada@neptune.kanazawa-it.ac.jp}}
\vspace*{0.015truein}
\centerline{\footnotesize\it Office for Promoting Co-Creation Education Administrative Office, Kanazawa Institute of Technology, Nonoichi, Ishikawa 921-8501, Japan}
\vspace*{10pt}
\vspace*{0.225truein}

\vspace*{0.21truein}

\begin{abstract}

We focus on the three-state quantum walk(QW) in one dimension.
In this paper, we give the stationary measure in general condition, originated from the eigenvalue problem.
Firstly, we get the transfer matrices by our new recipe, and solve the eigenvalue problem. Then we obtain the general form of the stationary measure for concrete initial state and eigenvalue. We also show some specific examples of the stationary measure for the three-state QW. 
One of the interesting and crucial future problems is to make clear the whole picture of the set of stationary measures. 

\end{abstract}

\vspace*{10pt}
\vspace*{3pt}

\vspace*{1pt}\textlineskip %) USE THIS MEASUREMENT WHEN THERE IS
   %) A SECTION HEADING
%\vspace*{-0.5pt}
%\noindent
%%%%%%%%%%%%%%%%%%%%%%%%%%%%%%%%
%put the text of the paper here
%%%%%%%%%%%%%%%%%%%%%%%%%%%%%%%%
\section{Introduction} 
\label{Introduction}
Owing to their specific properties, quantum walks(QWs) have attracted much attention in various fields, such as quantum algorithms \cite{ambainisEtAl2013,kempeEtAl2003}, and topological insulators \cite{kitagawaEtAl2010}. For the rich application of QWs, it is important to further study
both analytically and numerically. Indeed, over the past decade many
researchers have investigated the asymptotic behaviors of QWs from various viewpoints  
 \cite{choho2013,joyemerkli2010,konnoweak2005,konnoluczaksegawa2013,konnoyoo2013,wojcik2012}. \\
\indent
So far, localization and the ballistic spreading have been known as the characteristic properties of QWs     
 \cite{scholz,schudoEtAl2004,inuiEtAl2003,joyemerkli2010}. They are described mathematically by two kinds of limit theorems, which are composed of measures, i.e., the time-averaged limit measure corresponding to localization, and the weak limit measures describing the ballistic spreading  \cite{konnoluczaksegawa2013}. We should note that the weak limit measure is consisted by the Dirac measure part corresponding to localization and absolutely continuous part, expressing the ballistic spreading.
The weak convergence theorem for various types of QWs in one dimension, such as Hadamard walk  
 \cite{konnoweak2005}, Grover walk \cite{konno2008}, the two-phase QWs \cite{endoEtAl2016, endobusekonno2015} were derived. \\
\indent
Recently, the stationary measure of the QW has received peculiar interests as another key measure for the distribution of QW.  
The stationary measure provides the stationary distribution, for instance. Here we briefly review the past studies of the stationary measure. Comparing to the study of the stationary distributions of Markov chains, the
corresponding study of QWs has not been done sufficiently. 
Most of the present papers deal with stationary measures of the discrete-time case mainly on $\mathbb{Z}$, where $\mathbb{Z}$ is the set of integers. 
As for the stationary measure of the two-state QW,
Konno et al. \cite{konnoluczaksegawa2013} treated QW with a single defect at the origin and showed that a stationary measure with exponential decay for the position is identical to the time-averaged
limit measure for the QW.  \cite{konno2014} investigated stationary
measures for various types of QWs. Endo et al. \cite{endoEtAl2014} got a stationary measure of the QW with a single defect whose quantum coins are defined by the Hadamard matrix at $x\neq0$ and the rotation matrix at $x = 0$. Endo and Konno \cite{endokonno2014} calculated a stationary measure of QW with a single defect which was introduced and studied
by Wojcik et al . \cite{wojcik2012}. Furthermore, Endo et al. \cite{endobusekonno2015} and Endo et al.   
 \cite{endoEtAl2015} obtained stationary measures of the
two-phase QWs without defect and with a single defect, respectively, and investigated the relation to localization and the topological insulator. 
Konno and Takei \cite{konnotakei2015} considered stationary
measures of QWs and provided non-uniform stationary measures of non-diagonal QWs. They also showed that the set of the stationary
measures generally contains uniform measure. As for the stationary measure of the three-state QW,
Konno \cite{konno2014} obtained the stationary measures of the three-state Grover walk. Moreover, Wang et al. \cite{wan2015} investigated stationary measures of the three-state Grover walk with a single defect at the origin. Kawai et al. \cite{kawaiEtAl2017} constructed stationary measures for some types of the three-state QWs by using the reduction matrices. As for higher dimensional case, Komatsu and Konno \cite{komatsukonno2017} gave the stationary measures of the Grover walk on $\mathbb{Z}^{d}(d\in\mathbb{N})$, where $\mathbb{N}$ is the set of natural numbers.
In this paper, we consider stationary measures for the three-state QWs.\\
\indent
Since there are rich applications of space-inhomogeneous QWs, such as the applications for networks \cite{sadowskiEtAl2016} and topological insulator \cite{endobusekonno2015}, the research on the mathematical aspects of space-inhomogeneous QW to  exactly grasp the asymptotic behavior is important topic in the theoretical study of QWs \cite{ahlbrechtEtAl2011,canteroEtAl2012,konno2009,konno2010}. However, comparing to the study on the two-state QW, the three-state QW has not been treated enough. As for the stationary measures, there are few results for the quantum systems in comparison with that of the classical systems.
In particular, the stationary measures for the classical systems have been known to be constant or exponential increase:
\[\mu(x)=\left\{ \begin{array}{ll}
c & (\in l^{\infty}\!-\!space \;on\; \mathbb{Z}), \\
\\
\left(\dfrac{q}{p}\right)^{x} & (\notin l^{\infty}\!-\!space \;on\; \mathbb{Z}), \\
\end{array} \right.\] 
with $c,p,q\in\mathbb{R}_{\geq0}$, and $0\neq p<q,\;p+q=1$. Here $\mathbb{R}_{\geq0}$ is the set of the real numbers $r\geq0$. Also, in the study of the generalized eigenfunctions in the Spectral scattering theory, the eigenfunctions in $l^{\infty}$-space have been actively discussed \cite{morioka2019}. According to the background, it is very important to study the stationary measures that are not in $l ^ 2$-space on $\mathbb{Z}$.\\
\indent
Our main result is the general form of the stationary measure obtained by the transfer matrices of the three-state QW.\\
\indent
The rest of this paper is organized as follows. In Section \ref{main},
we give the definition of the three-state QW which is the main target in this paper, and present our main result. 
We show specific examples of our results in Section \ref{discuss}. 
In Appendix, we give the proof of Theorem \ref{transfer}.
 
%%%%%%%%%%%%%%%%%%%%%%%%%%%%%%%%%%%%%%%%%%%%%%%%%%%%%%%%%%%%%%%%%%%%%%%%%%
%(Model & results)%
%%%%%%%%%%%%%%%%%%%%%%%%%%%%%%%%%%%%%%%%%%%%%%%%%%%%%%%%%%%%%%%%%%%%%%%%%%
\section{Stationary measure in general condition originated from the eigenvalue problem}
\label{main}
\noindent
First, we introduce discrete-time space-inhomogeneous QW with three-state on $\mathbb{Z}$, 
which is a quantum version of classical random walk with an additional coin state. 
The particle has a coin state at time $n$ and position $x$ described by a three dimensional vector in $\mathbb{C}^{3},$ where $\mathbb{C}$ is the set of complex numbers: 
\[\Psi_{n}(x)=
\begin{bmatrix}
\Psi^{L}_{n}(x)\\
\Psi^{O}_{n}(x)\\
\Psi^{R}_{n}(x)
\end{bmatrix}\quad(x\in\mathbb{Z}).\]
The upper and lower elements express the left and right chiralities, respectively, and the middle element corresponds to the loop.
The time evolution is determined by $3\times 3$ unitary matrices $U_{x}$:
\begin{align*}
U_{x}=\begin{bmatrix}
a_{x} & b_{x} & c_{x}\\
d_{x} & e_{x} & f_{x} \\
g_{x} & h_{x} & i_{x}
\end{bmatrix},
\end{align*}
where $x\in\mathbb{Z},$ and $a_{x}, b_{x}, c_{x}, d_{x}, e_{x}, f_{x},
g_{x}, h_{x}, i_{x}\in\mathbb{C}$.\\
Here the time evolution is determined by the recurrence formula
\begin{align*}
\Psi_{n+1} (x) = P_{x+1}\Psi_n (x+1) + R_{x}\Psi_n (x)+Q_{x-1}\Psi_n (x-1) \quad (x \in \mathbb{Z}),
\end{align*}
where
\begin{align*}
P_x =\begin{bmatrix} 
a_{x} & b_{x} & c_{x} \\ 
0 & 0 & 0\\
0 & 0 & 0\\
\end{bmatrix},
\qquad 
R_x =\begin{bmatrix} 
0 & 0 & 0 \\ 
d_{x} & e_{x} & f_{x} \\
0&  0 & 0 \\
\end{bmatrix},
\qquad 
Q_x = 
\begin{bmatrix} 
0 & 0 & 0\\ 
0 & 0 & 0\\
g_{x} & h_{x} & i_{x}
\end{bmatrix}, \\
\end{align*}
with $U_x = P_x + R_x +Q_x$. Note that $P_x$ and $Q_x$ correspond to the left and right movements, respectively, and $R_{x}$ represents the loop.
Here we remark that the walker steps dependent on position. \\
\indent
From now on, we introduce the stationary measure of the QW. Let
\[\Psi_{n}= \left[\cdots,\begin{bmatrix}
\Psi_{n}^{L}(-1)\\
\Psi_{n}^{O}(-1)\\
\Psi_{n}^{R}(-1)\end{bmatrix},\begin{bmatrix}
\Psi_{n}^{L}(0)\\
\Psi_{n}^{O}(0)\\
\Psi_{n}^{R}(0)\end{bmatrix},\begin{bmatrix}
\Psi_{n}^{L}(1)\\
\Psi_{n}^{O}(1)\\
\Psi_{n}^{R}(1)\end{bmatrix},\cdots\right]{}^T\!\in(\mathbb{C}^{3})^{\mathbb{Z}},\]
and
\[U^{(s)}=\begin{bmatrix}
\ddots&\vdots&\vdots&\vdots&\vdots&\vdots&\iddots\\
\cdots&R_{-2}&P_{-1}&O&O&O&\cdots\\
\cdots&Q_{-2}&R_{-1}&P_{0}&O&O&\cdots\\
\cdots&O&Q_{-1}&R_{0}&P_{1}&O&\cdots\\
\cdots&O&O&Q_{0}&R_{1}&P_{2}&\cdots\\
\cdots&O&O&O&Q_{1}&R_{2}&\cdots\\
\iddots&\vdots&\vdots&\vdots&\vdots&\vdots&\ddots
\end{bmatrix}\;\;\;
with\;\;\;O=\begin{bmatrix}0&0&0\\0&0&0\\0&0&0\end{bmatrix},\]
where $T$ means the transposed operation. Then the state of the QW at time $n$ is given by
$\Psi_{n}=(U^{(s)})^{n}\Psi_{0}$ for any $n\geq0$. 
Let $\mathbb{R}_{+}=[0,\infty)$. Here we introduce a map 
$\phi:(\mathbb{C}^{3})^{\mathbb{Z}}\rightarrow \mathbb{R}_{+}^{\mathbb{Z}}$
such that for $\Psi_{n},$
we put
\[\phi(\Psi)(x) = |\Psi^{L}(x)|^{2} +|\Psi^{O}(x)|^{2}+ |\Psi^{R}(x)|^{2}\;\;\;(x\in\mathbb{Z}).\]
Then we define the measure $\mu:\mathbb{Z}\to\mathbb{R}_{+}$ by
\[\mu(x)=\phi(\Psi)(x).\]
We should note that $\mu(x)$ is a measure of the QW at position $x$.
Now put 
\begin{align*}
\Sigma_{s}=\{\phi(\Psi_{0})\in\mathbb{R}_{+}^{\mathbb{Z}}: \exists\;\Psi_{0},
\;{\rm s.t.}\;\phi((U^{(s)})^{n}\Psi_{0})=\phi(\Psi_{0})\;\forall n\geq 0\},\end{align*}
and we call the element of $\Sigma_{s}$, {\it the stationary measure} of the QW.\\
Here let $S^{1}=\{z\in\mathbb{C}:|z|=1\},$ and consider the eigenvalue problem $U^{(s)}\Psi=\lambda\Psi(\Psi\in {\rm Map}(\mathbb{Z},\mathbb{C}^{3}),\lambda\in S^{1})$.
Since $U^{(s)}$ is unitary, we directly see $\phi(U^{(s)}\Psi)\in\Sigma_{s}$.\\
\indent
Now we solve the eigenvalue problem as follows. The proof is devoted to Appendix.
\par\noindent
\begin{theorem}
\label{transfer}
Let $\{U_{y}\}_{y\in\mathbb{Z}}$ be the set of $y$-parameterized unitary matrices of the three-state inhomogeneous QW, 
and $\Psi(x)=[\Psi^{L}(x),\;\Psi^{O}(x),\;\Psi^{R}(x)]{}^T\!$ be the probability amplitude. Note that there is a restriction for the initial state $\Psi(0)$ \cite{endokawaikonno2016}.
Then the solutions for $U^{(s)}\Psi=\lambda\Psi(\Psi\in {\rm Map}(\mathbb{Z},\mathbb{C}^{3}),\lambda\in S^{1})$
are
\begin{eqnarray*}
\Psi(x)=
\left\{\begin{array}{ll}
\prod^{x}_{y=1}T^{(+)}_{y}\Psi(0) & (x\geq 1), \\\
\Psi(0)& (x=0), \\\
\prod^{x}_{y=-1}T^{(-)}_{y}\Psi(0)  & (x\leq -1), 
\end{array}\right.
\end{eqnarray*} 
where $T^{(\pm)}_{y}$ are the transfer matrices defined by 
\begin{align*}
T_{y}^{(+)}&=\begin{bmatrix}t^{(+)}_{11}&t^{(+)}_{12}&t^{(+)}_{13}\\ 
t^{(+)}_{21}&t^{(+)}_{22}&t^{(+)}_{23}\\
t^{(+)}_{31}&t^{(+)}_{32}&t^{(+)}_{33} 
 \end{bmatrix},\quad T^{(-)}_{y}=\begin{bmatrix}t^{(-)}_{11} &t^{(-)}_{12}& t^{(-)}_{13}\\ 
t^{(-)}_{21}&t^{(-)}_{22} &t^{(-)}_{23}\\
t^{(-)}_{31}&t^{(+)}_{32}&t^{(+)}_{33}
\end{bmatrix},\end{align*} \\
with
\begin{align*}
t^{(+)}_{11}&=\dfrac{(\lambda-e_{y})(\lambda^{2}-g_{y-1}c_{y})-g_{y-1}b_{y}f_{y}}{\lambda\{a_{y}(\lambda-e_{y})+b_{y}d_{y}\}},\;
t^{(+)}_{12}= -\dfrac{h_{y-1}\{b_{y}f_{y}+c_{y}(\lambda-e_{y})\}}{\lambda\{a_{y}(\lambda-e_{y})+b_{y}d_{y}\}} \\
t^{(+)}_{13}&=-\dfrac{i_{y-1}\{b_{y}f_{y}+c_{y}(\lambda-e_{y})\}}{\lambda\{a_{y}(\lambda-e_{y})+b_{y}d_{y}\}},\;
t^{(+)}_{21}=\dfrac{\lambda^{2}d_{y}+g_{y-1}(a_{y}f_{y}-c_{y}d_{y})}{\lambda\{a_{y}(\lambda-e_{y})+b_{y}d_{y}\}}\\
t^{(+)}_{22}&= \dfrac{h_{y-1}(a_{y}f_{y}-c_{y}d_{y})}{\lambda\{a_{y}(\lambda-e_{y})+b_{y}d_{y}\}},\;
t^{(+)}_{23}= \dfrac{i_{y-1}(a_{y}f_{y}-c_{y}d_{y})}{\lambda\{a_{y}(\lambda-e_{y})+b_{y}d_{y}\}}\\
t^{(+)}_{31}&=\dfrac{g_{y-1}}{\lambda},\;t^{(+)}_{32}=\dfrac{h_{y-1}}{\lambda},\;t^{(+)}_{33}=\dfrac{i_{y-1}}{\lambda},
\end{align*}
and
\begin{align*}
t^{(-)}_{11}&=\dfrac{a_{y+1}}{\lambda},\;
t^{(-)}_{12}=\dfrac{b_{y+1}}{\lambda},\;
t^{(-)}_{13}=\dfrac{c_{y+1}}{\lambda},\;\\
t^{(-)}_{21}&=-\dfrac{a_{y+1}(f_{y}g_{y}-i_{y}d_{y})}{\lambda\{h_{y}f_{y}+i_{y}(\lambda-e_{y})\}},\;
t^{(-)}_{22}= -\dfrac{b_{y+1}(f_{y}g_{y}-i_{y}d_{y})}{\lambda\{h_{y}f_{y}+i_{y}(\lambda-e_{y})\}},\\
t^{(-)}_{23}&=\dfrac{\lambda^{2}f_{y}-c_{y+1}(f_{y}g_{y}-i_{y}d_{y})}{\lambda\{h_{y}f_{y}+i_{y}(\lambda-e_{y})\}},\;
t^{(-)}_{31}=-\dfrac{a_{y+1}\{h_{y}d_{y}+g_{y}(\lambda-e_{y})\}}{\lambda\{h_{y}f_{y}+i_{y}(\lambda-e_{y})\}},\\
t^{(-)}_{32}&=-\dfrac{b_{y+1}\{h_{y}d_{y}+g_{y}(\lambda-e_{y})\}}{\lambda\{h_{y}f_{y}+i_{y}(\lambda-e_{y})\}},\;t^{(-)}_{33}=\dfrac{(\lambda-e_{y})(\lambda^{2}-g_{y}c_{y+1})-h_{y}c_{y+1}d_{y}}{\lambda\{h_{y}f_{y}+i_{y}(\lambda-e_{y})\}}.
\end{align*}

Furthermore, the stationary measure is given by $\mu(x)=\phi(\Psi)(x)=\|\Psi(x)\|^{2}\;(x\in\mathbb{Z})$.
\end{theorem}
We note that the initial state develops by the transfer matrices. 
We can obtain the stationary measure for given three-state QW and initial state by calculating the transfer matrices. Also, we should remark that we cannot apply Theorem \ref{transfer} in the case that the elements of the transfer matrices diverge.
Our result is more effective for the three-state QWs in general than the result obtained by Kawai et al. \cite{kawaiEtAl2017}, since they use the reduced matrix, which restricts the models that can be analyzed.

\setcounter{footnote}{0}
\renewcommand{\thefootnote}{\alph{footnote}}

%%%%%%%%%%%%%%%%%%%%%%%%%%%%%%%%%%%%%%%%%%%%%%%%%%%%%%%%%%%%%%%%%%%%%%%%%
%(Examples)%
%%%%%%%%%%%%%%%%%%%%%%%%%%%%%%%%%%%%%%%%%%%%%%%%%%%%%%%%%%%%%%%%%%%%%%%%%
\section{Examples}
\label{discuss}
%先行研究との比較を載せる%
In this section, we exhibit some concrete typical examples of our result, Theorem \ref{transfer}.

\begin{enumerate}
\item The Grover walk:\\
First, we consider the Grover walk defined by the unitary matrix
\begin{align*}
U_{x}=\dfrac{1}{3}\begin{bmatrix}-1&2&2\\2&-1&2\\2&2&-1\end{bmatrix}.\end{align*}
\noindent

We got the stationary amplitude originated from the eigenvalue problem $U^{(s)}\Psi=\lambda\Psi$ as follows.
\begin{proposition}
Let $\{M_{y}\}_{y\in\mathbb{Z}}$ be the set of $y\in\mathbb{Z}$-parameterized unitary matrices of the Grover walk, 
and $\Psi(x)=[\Psi^{L}(x),\;\Psi^{O}(x),\;\Psi^{R}(x)]{}^T$ be the probability amplitude. 
Put $\alpha=\Psi^{L}(0),\;\gamma=\Psi^{O}(0)$ and $\beta=\Psi^{R}(0)$.
Then we have $(1+3\lambda)\Psi^{O}(0)=2\Psi^{L}(0)+2\Psi^{R}(0)$, and for every $\lambda\in S^{1}$, we can choose $\Psi(0)=[\alpha,0,-\alpha]{}^T$ as an initial state. Now we take $\lambda=-1$, and the solutions for $U^{(s)}\Psi=\lambda\Psi(\Psi\in Map(\mathbb{Z},\mathbb{C}^{2}),\lambda\in S^{1}),$
are
\begin{eqnarray*}
\Psi(x)=
\left\{\begin{array}{ll}
\prod^{x}_{y=1}T^{(+)}_{y}\Psi(0) & (x\geq 1), \\\
\Psi(0)& (x=0), \\\
\prod^{x}_{y=-1}T^{(-)}_{y}\Psi(0)  & (x\leq -1), 
\end{array}\right.
\end{eqnarray*} 
where $T^{(\pm)}_{y}$ are the transfer matrices defined by 

\begin{align*}
T_{y}^{(+)}=\begin{bmatrix}
1&0&0\\
\\Konno
-\dfrac{1}{3}&\dfrac{2}{3}&-\dfrac{1}{3}\\
\\
-\dfrac{2}{3}&-\dfrac{2}{3}&\dfrac{1}{3} \end{bmatrix}, \quad
T^{(-)}_{y}=\begin{bmatrix}
\dfrac{1}{3}&-\dfrac{2}{3}&-\dfrac{2}{3}\\
\\
-\dfrac{1}{3}&\dfrac{2}{3}&-\dfrac{1}{3}\\
\\
0&0&1
\end{bmatrix}.\\
\end{align*}
\end{proposition}
For $\Psi(0)=[\alpha,0,-\alpha]{}^T$, we see
\begin{align*}
\Psi(x)=[\alpha,0,-\alpha]{}^T\quad(x\in\mathbb{Z}),
\end{align*}
and therefore, we obtain
\begin{align*}
\mu(x)=
2|\alpha|^{2}\quad (x\in\mathbb{Z}).
\end{align*}
We see that the stationary measure is uniform, which is same with the result in Section $4$ of \cite{kawaiEtAl2017}.\\
%%%%%%%%%%%%%%%%%%%%%%%%%%%%%%%%%%%%%%%%%%%%%%%%%%%%%%%%%%%%%%%%%%%%%%%%%%%%%%%%%%%%%%%%%%%
\item  The Grover walk+1-defect model:\\
Next, we focus on the QW defined by the unitary matrices
\begin{align*}
U_{x}=\left\{ \begin{array}{ll}
\dfrac{1}{3}\begin{bmatrix}-1&2&2\\2&-1&2\\2&2&-1\end{bmatrix}\quad (x\neq0), \\
\\
\dfrac{\rho}{3}\begin{bmatrix}-1&2&2\\2&-1&2\\2&2&-1 \end{bmatrix}\quad (x=0),
\end{array} \right.\end{align*}
with $\rho=e^{i\phi}\;(\phi\in[0,\infty))$.\\
The QW is given by putting the weight $\rho$ at $x=0$ to the Grover walk.
We derived the stationary amplitude of the eigenvalue problem $U^{(s)}\Psi=\lambda\Psi$ as follows.

\begin{proposition}
Let $\{M_{y}\}_{y\in\mathbb{Z}}$ be the set of $y\in\mathbb{Z}$-parameterized unitary matrices of the Grover walk+1-defect model, 
and $\Psi(x)=[\Psi^{L}(x),\;\Psi^{O}(x),\;\Psi^{R}(x)]{}^T$ be the probability amplitude. 
Put $\alpha=\Psi^{L}(0),\;\gamma=\Psi^{O}(0)$ and $\beta=\Psi^{R}(0)$.
Then we see $(1-3\lambda)\Psi^{O}(0)=2\Psi^{L}(0)+2\Psi^{R}(0)$, and for every $\lambda\in S^{1}$, we can take $\Psi(0)=[\alpha,0,-\alpha]{}^T$ as an initial state. Now we take $\lambda=-1$, and the solutions for $U^{(s)}\Psi=\lambda\Psi(\Psi\in Map(\mathbb{Z},\mathbb{C}^{2}),\lambda\in S^{1}),$
are
\begin{eqnarray*}
\Psi(x)=
\left\{\begin{array}{ll}
\prod^{x}_{y=1}T^{(+)}_{y}\Psi(0) & (x\geq 1), \\\
\Psi(0)& (x=0), \\\
\prod^{x}_{y=-1}T^{(-)}_{y}\Psi(0)  & (x\leq -1), 
\end{array}\right.
\end{eqnarray*} 
where $T^{(\pm)}_{y}$ are the transfer matrices defined by 

\begin{align*}
T_{1}^{(+)}=\begin{bmatrix}
1&0&0\\
\\
-\dfrac{5}{3}&-\dfrac{2}{3}&\dfrac{1}{3}\\
\\
\dfrac{2}{3}&\dfrac{2}{3}&-\dfrac{1}{3} \end{bmatrix}, \quad
T^{(-)}_{-1}=\begin{bmatrix}
-\dfrac{1}{3}&\dfrac{2}{3}&\dfrac{2}{3}\\
\\
\dfrac{1}{3}&-\dfrac{2}{3}&-\dfrac{5}{3}\\
\\
0&0&1
\end{bmatrix},\\
T_{y}^{(+)}=\begin{bmatrix}
1&0&0\\
\\
-\dfrac{1}{3}&\dfrac{2}{3}&-\dfrac{1}{3}\\
\\
-\dfrac{2}{3}&-\dfrac{2}{3}&\dfrac{1}{3} \end{bmatrix}, \quad
T^{(-)}_{y}=\begin{bmatrix}
\dfrac{1}{3}&-\dfrac{2}{3}&-\dfrac{2}{3}\\
\\
-\dfrac{1}{3}&\dfrac{2}{3}&-\dfrac{1}{3}\\
\\
0&0&1
\end{bmatrix}\quad(|y|\geq2).\\
\end{align*}
\end{proposition}

For $\Psi(0) =[\alpha,0,-\alpha]{}^T,$ we obtain
\begin{align*}
\Psi(x)=\left\{ \begin{array}{ll}
\left[\alpha,0,-\alpha\right]{}^T\!&(|x|\neq 1),\\
\\
\left[\alpha,-2\alpha,\alpha\right]{}^T\!&(x=1),\\
\\
\left[-\alpha,2\alpha,-\alpha\right]{}^T\!&(x=-1).
\end{array} \right.
\end{align*}
Hence, we have
\begin{align*}\mu(x)=2|\alpha|^{2}
\left\{ \begin{array}{ll}
1 & (|x|\neq 1), \\
\\
3 & (|x|=1). \\
\end{array} \right.
\end{align*}
Thereby, the stationary measure is not uniform, and the result does not coincide with the result in Section $4$ of \cite{endokawaikonno2016}. The defect at the origin seems to influence the value of the stationary measure not at the origin, but at both sides of the origin.\\
%%%%%%%%%%%%%%%%%%%%%%%%%%%%%%%%%%%%%%%%%%%%%%%%%%%%%%%%%%%%%%%%%%%%%%%%%%%%
\item The Fourier QW:
Here we treat the Fourier QW defined by the unitary matrices
\begin{align*}
U_{x}=
\dfrac{1}{\sqrt{3}}\begin{bmatrix}1&1&1\\1&\omega&\omega^{2}\\1&\omega^{2}&\omega\end{bmatrix} 
\end{align*}
with $\omega=e^{\frac{2}{3}i\pi}$.\\

By putting $\lambda=(1+2\omega)/\sqrt{3}$, we got the stationary amplitude of the eigenvalue problem $U^{(s)}\Psi=\lambda\Psi$ as follows.

\begin{proposition}
Let $\{M_{y}\}_{y\in\mathbb{Z}}$ be the set of $y\in\mathbb{Z}$-parameterized unitary matrices of the Fourier QW, 
and $\Psi(x)={}^T\![\Psi^{L}(x),\;\Psi^{O}(x),\;\Psi^{R}(x)]$ be the probability amplitude. 
Then we see $(\sqrt{3}\lambda-\omega)\Psi^{O}(0)=\Psi^{L}(0)+\omega^{2}\Psi^{R}(0)$, and for every $\lambda\in S^{1}$, we can choose $\Psi(0)=[\alpha,0,-\alpha\omega^{-2}]{}^T$ as an initial state. Now we take $\lambda=i$, and the solutions for $U^{(s)}\Psi=\lambda\Psi(\Psi\in Map(\mathbb{Z},\mathbb{C}^{2}),\lambda\in S^{1}),$
are
\begin{eqnarray*}
\Psi(x)=
\left\{\begin{array}{ll}
\prod^{x}_{y=1}T^{(+)}_{y}\Psi(0) & (x\geq 1), \\\
\Psi(0)& (x=0), \\\
\prod^{x}_{y=-1}T^{(-)}_{y}\Psi(0)  & (x\leq -1), 
\end{array}\right.
\end{eqnarray*} 
where $T^{(\pm)}_{y}$ are the transfer matrices defined by 

\begin{align*}
T_{y}^{(+)}=\begin{bmatrix}
\omega&0&0\\
\\
\dfrac{-\omega-5}{3\omega}&\dfrac{1-\omega}{3}&\dfrac{1-\omega}{3\omega}\\
\\
\dfrac{1}{\omega(1-\omega)}&\dfrac{\omega}{1-\omega}&\dfrac{1}{1-\omega} \end{bmatrix}, \quad
T^{(-)}_{y}=\begin{bmatrix}
\dfrac{1}{\omega(1-\omega)}&\dfrac{1}{\omega(1-\omega)}&\dfrac{1}{\omega(1-\omega)}\\
\\
\dfrac{-1-2\omega}{3(1+\omega)}&\dfrac{-1-2\omega}{3(1+\omega)}&\dfrac{-4-5\omega}{3(1+\omega)}\\
\\
0&0&1
\end{bmatrix}\quad(|y|\geq1).\\
\end{align*}
\end{proposition}
For $\Psi(0)={}^T\![\alpha,0,-\alpha\omega^{-2}]$, we get
\begin{align*}
\Psi(x)=\left\{ \begin{array}{ll}
{}^T\!\left[\alpha,0,-\alpha\omega^{-2}\right]&(x=3m),\\
\\
{}^T\!\left[\alpha\omega,\alpha\dfrac{-2-\omega}{\omega},-\alpha\omega^{-2}\right]&(x=3m+1),\\
\\
{}^T\!\left[\alpha\omega^{2},\alpha\dfrac{-2\omega-1}{\omega},-\alpha\omega^{-2}\right]&(x=3m+2),\\
\\
{}^T\!\left[\alpha\omega^{-1},\alpha\dfrac{-2\omega-1}{\omega},-\alpha\omega^{-2}\right]&(x=-3m-1),\\
\\
{}^T\!\left[\alpha\omega^{-2},\alpha\dfrac{-2-\omega}{\omega},-\alpha\omega^{-2}\right]&(x=-3m-2),
\end{array} \right.
\end{align*}
where $m\in\mathbb{Z}_{\geq 0}$ with $\mathbb{Z}_{\geq0}=\{0,1,2,\cdots\}$.
Hence, we obtain
\begin{align*}
\mu(x)=|\alpha|^{2}\times\left\{ \begin{array}{ll}
2&(x=3m),\\
\\
5&(x\neq3m).\\
\end{array} \right.
\end{align*}
We notice that the stationary measure has period $3$, which is same with the result in Section $4$ of \cite{kawaiEtAl2017}.
\end{enumerate}

%%%%%%%%%%%%%%%%%%%%%%%%%%%%%%%%%%%%%%%%%%%%%%%%%%%%%%%%%%%%%%%%%%%%%%%%%
%(Summary)%
%%%%%%%%%%%%%%%%%%%%%%%%%%%%%%%%%%%%%%%%%%%%%%%%%%%%%%%%%%%%%%%%%%%%%%%%%
\section{Summary}
In this paper, we obtained for the three-state QW, the general form of the stationary measure originated from the eigenvalue problem.
Our method is mainly based on the transfer matrices, and is more effective than the reduction method developed by Kawai et al. \cite{kawaiEtAl2017}, since we do not need to reduct to two-state model in our recipe.
By using our result, we can derive various types of stationary measures for many types of the three-state QWs, which contributes to clear the whole picture of the set of stationary measures.
One of our future problems is to examine the eigenvalues and eigenspaces, which directly connects to Spectral theory.
Also, to investigate the stationary measure which does not come from the eigenvalue problem is fundamental for the study of the stationary measure of the QW.
On the other hand, to explore the relation between the three-state QWs and physical phenomenon, such as the topological insulator, leads to the application of the three-state QWs to quantum information sciences.

%%%%%%%%%%%%%%%%%%%%%%%%%%%%%%%%%%%%%%%%%%%%%%%%%%%%%%%%%%%%%%%%%%%%%%%%%%
%(acknowledgement)%
%%%%%%%%%%%%%%%%%%%%%%%%%%%%%%%%%%%%%%%%%%%%%%%%%%%%%%%%%%%%%%%%%%%%%%%%%%

\setcounter{footnote}{0}
\renewcommand{\thefootnote}{\alph{footnote}}
\section*{Acknowledgments}
TE is supported by financial support of Postdoctoral Fellowship from Japan Society for the
Promotion of Science.

\nonumsection{References}

%%%%%%%%%%%%%%%%%%%%%%%%%%%%%%%%%%%%%%%%%%%%%%%%%%%%%%%%%%%%%%%%%%%%%%%%%%
%(Appendix A)%
%%%%%%%%%%%%%%%%%%%%%%%%%%%%%%%%%%%%%%%%%%%%%%%%%%%%%%%%%%%%%%%%%%%%%%%%%%

\nonumsection{Appendix}
\label{prooftheo}
In Appendix, we derive the transfer matrices $T^{(\pm)}_{x}$ of the three-state QW, which leads to Theorem \ref{transfer}.
Since $\lambda\Psi=U^{(s)}\Psi(\Psi\in {\rm Map}(\mathbb{Z},\mathbb{C}^{3}),\lambda\in S^{1})$ holds, we have
\begin{align}
\lambda
\Psi^{L}(x)&=a_{x+1}\Psi^{L}(x+1)+b_{x+1}\Psi^{O}(x+1)+c_{x+1}\Psi^{R}(x+1),\label{peq1}\\
\lambda\Psi^{O}(x)&=d_{x}\Psi^{L}(x)+e_{x}\Psi^{O}(x)+f_{x}\Psi^{R}(x),\label{peq2}\\
\lambda\Psi^{R}(x)&=g_{x-1}\Psi^{L}(x-1)+h_{x-1}\Psi^{O}(x-1)+i_{x-1}\Psi^{R}(x-1)\label{peq3}.
\end{align}
\begin{enumerate}
\item Case of $T^{(-)}_{x}$: By using Eq.\eqref{peq1}, we can write down
\begin{align*}
\begin{bmatrix}
\Psi^{L}(x)\\
\Psi^{O}(x)\\
\Psi^{R}(x)\\
\end{bmatrix}=
\begin{bmatrix}
\dfrac{a_{x+1}}{\lambda} \dfrac{b_{x+1}}{\lambda} \dfrac{c_{x+1}}{\lambda}\\
(A)\\
(B)\\
\end{bmatrix}\begin{bmatrix}
\Psi^{L}(x+1)\\
\Psi^{O}(x+1)\\
\Psi^{R}(x+1)\\
\end{bmatrix}.
\end{align*}
From now on, we derive $(A)$ and $(B)$, which directly leads to $T^{(-)}_{x}$.
\begin{itemize}
\item For (A):
Owing to Eqs.\eqref{peq3} and \eqref{peq2}, we have
\begin{align}
h_{x}\Psi^{O}(x)=\lambda\Psi^{R}(x+1)-g_{x}\Psi^{L}(x)-i_{x}\Psi^{R}(x),\label{peq4}
\end{align}
and
\begin{align}
\Psi^{R}(x)=\dfrac{\lambda}{f_{x}}\Psi^{O}(x)-\dfrac{d_{x}}{f_{x}}\Psi^{L}(x)-\dfrac{e_{x}}{f_{x}}\Psi^{O}(x),\label{peq5}
\end{align}
respectively.
Substituting Eq.\eqref{peq5} into Eq.\eqref{peq4}, we get
\begin{align*}
\left\{h_{x}+\dfrac{i_{x}}{f_{x}}(\lambda-e_{x})\right\}\Psi^{O}(x)=\lambda\Psi^{R}(x+1)-\left(g_{x}-\dfrac{i_{x}d_{x}}{f_{x}}\right)\Psi^{L}(x).\label{peq6}
\end{align*}
Taking into account of Eq.\eqref{peq1}, we obtain
\begin{align}
\Psi^{O}(x)=-\dfrac{a_{x+1}(f_{x}g_{x}-i_{x}d_{x})}{\lambda\{h_{x}f_{x}+i_{x}(\lambda-e_{x})\}}\Psi^{L}(x+1)-\dfrac{b_{x+1}(f_{x}g_{x}-i_{x}d_{x})}{\lambda\{h_{x}f_{x}+i_{x}(\lambda-e_{x})\}}\Psi^{O}(x+1)\nonumber\\
+\dfrac{\lambda^{2}f_{x}-c_{x+1}(f_{x}g_{x}-i_{x}d_{x})}{\lambda\{h_{x}f_{x}+i_{x}(\lambda-e_{x})\}}\Psi^{R}(x+1),
\end{align}
which implies (A).

\item For (B):
From Eq.\eqref{peq3}, we get
\begin{align}
i_{x}\Psi^{R}(x)=\lambda\Psi^{R}(x+1)-g_{x}\Psi^{L}(x)-h_{x}\Psi^{O}(x).\label{peq6}
\end{align}
Substituting Eq.(16) and Eq.\eqref{peq1} into Eq.\eqref{peq6}, we see
\begin{align*}
\Psi^{R}(x)=-\dfrac{a_{x+1}\{h_{x}d_{x}+g_{x}(\lambda-e_{x})\}}{\lambda\{h_{x}f_{x}+i_{x}(\lambda-e_{x})\}}\Psi^{L}(x+1)-\dfrac{b_{x+1}\{h_{x}d_{x}+g_{x}(\lambda-e_{x})\}}{\lambda\{h_{x}f_{x}+i_{x}(\lambda-e_{x})\}}\Psi^{O}(x+1)+\\\dfrac{(\lambda-e_{x})(\lambda^{2}-g_{x}c_{x+1})-h_{x}c_{x+1}d_{x}}{\lambda\{h_{x}f_{x}+i_{x}(\lambda-e_{x})\}}\Psi^{R}(x+1),
\end{align*}
which leads to (B).

\end{itemize}
\item Case of $T^{(+)}_{x}$: From, Eq.\eqref{peq3}, we can write down

\begin{align}
\begin{bmatrix}
\Psi^{L}(x)\\
\Psi^{O}(x)\\
\Psi^{R}(x)\\
\end{bmatrix}=
\begin{bmatrix}
(C)\\
(D)\\
\dfrac{g_{x-1}}{\lambda} \dfrac{h_{x-1}}{\lambda} \dfrac{i_{x-1}}{\lambda}\\
\end{bmatrix}\begin{bmatrix}
\Psi^{L}(x-1)\\
\Psi^{O}(x-1)\\
\Psi^{R}(x-1)\\
\end{bmatrix}.
\end{align}
Hereafter, we calculate $(C)$ and $(D)$, which contributes to $T^{(+)}_{x}$.
\begin{itemize}
\item For (C): By Eqs.\eqref{peq1} and \eqref{peq2}, we have
\begin{align}
a_{x}\Psi^{L}(x)=\lambda\Psi^{L}(x-1)-b_{x}\Psi^{O}(x)-c_{x}\Psi^{R}(x), \label{peq7}
\end{align}
and
\begin{align}
\Psi^{O}(x)=\dfrac{d_{x}}{\lambda-e_{x}}\Psi^{L}(x)+\dfrac{f_{x}}{\lambda-e_{x}}\Psi^{R}(x),\label{peq8}
\end{align}
respectively.
Substituting Eq.\eqref{peq8} into Eq.\eqref{peq7}, and taking into account of Eq.\eqref{peq3}, we obtain
\begin{align}
\Psi^{L}(x)=\dfrac{(\lambda-e_{x})(\lambda^{2}-g_{x-1}c_{x})-g_{x-1}b_{x}f_{x}}{\lambda\{a_{x}(\lambda-e_{x})+b_{x}d_{x}\}}\Psi^{L}(x-1)-\dfrac{h_{x-1}\{b_{x}f_{x}+c_{x}(\lambda-e_{x})\}}{\lambda\{a_{x}(\lambda-e_{x})+b_{x}d_{x}\}}\Psi^{O}(x-1)\nonumber\\
-\dfrac{i_{x-1}\{b_{x}f_{x}+c_{x}(\lambda-e_{x})\}}{\lambda\{a_{x}(\lambda-e_{x})+b_{x}d_{x}\}}\Psi^{R}(x-1),
\end{align}
which implies (C).

\item For (D):
Eq.\eqref{peq1}gives 
\begin{align}
b_{x}\Psi^{O}(x)=\lambda\Psi^{L}(x-1)-a_{x}\Psi^{L}(x)-c_{x}\Psi^{R}(x).\label{peq9}
\end{align}
Substituting Eq.(21) and Eq.\eqref{peq3} into Eq.\eqref{peq9}, we acquire
\begin{align*}
\Psi^{O}(x)=\dfrac{\lambda^{2}d_{x}+g_{x-1}(a_{x}f_{x}-c_{x}d_{x})}{\lambda\{a_{x}(\lambda-e_{x})+b_{x}d_{x}\}}\Psi^{L}(x-1)+\dfrac{h_{x-1}(a_{x}f_{x}-c_{x}d_{x})}{\lambda\{a_{x}(\lambda-e_{x})+b_{x}d_{x}\}}\Psi^{O}(x-1)\\
+\dfrac{i_{x-1}(a_{x}f_{x}-c_{x}d_{x})}{\lambda\{a_{x}(\lambda-e_{x})+b_{x}d_{x}\}}\Psi^{R}(x-1),
\end{align*}
which leads to (D).
\end{itemize}
\end{enumerate}
The above discussion completes the proof.

\begin{thebibliography}{99}
\bibitem{ahlbrechtEtAl2011}
A. Ahlbrecht, A. Alberti, D. Meschede, V. B. Scholz, A. H. Werner, and
	R. F. Werner: Bound Molecules in an Interacting Quantum Walk, arXiv1105.1051., 1-9 (2011)

\bibitem{scholz} 
A. Ahlbrecht, V. B. Scholz, and A. H. Werner: Disordered quantum walks in one lattice dimension, Journal of Mathematical Physics, {\bf 52}, 102201 (2011)
 
\bibitem{ambainisEtAl2013} 
A. Ambainis, A. Backurs, N. Nahimovs, R. Ozols, and A. Rivosh: 
Search by quantum walks on two-dimensional grid without amplitude amplification, Springer, Lecture Notes in Computer Science, {\bf 7582}, 87-97 (2013)
 
%\bibitem{burkovbalents} 
%A. A. Burkov and L. Balents:
% Weyl semimetal in a topological insulator multilayer, Physical Review Letters, {\bf 107}, 127205 (2011)
 
\bibitem{canteroEtAl2012} 
 M. J. Cantero, F. A. Grunbaum, L. Moral, and L. Velazquez:
 One-dimensional quantum walks with one defect, Reviews in Mathematical Physics, {\bf 24}, 1250002 (2012)
 
%\bibitem{chenEtAl2009}
% Y. L. Chen, J. G. Analytis, J.-H. Chu, Z. K. Liu, S.-K. Mo, X. L. Qi, H. J. Zhang, D. H. Lu, X. Dai, Z. Fang, S. C. Zhang, I. R. Fisher, Z. Hussain and Z.-X. Shen:
% Experimental realization of a three-dimensional topological insulator, Bi 2Te3, Science, {\bf 325}, 178-181 (2009) 

\bibitem{choho2013}
C.-I Chou and C.-L Ho:
Localization and recurrence of quantum walk in periodic potential on a line, Chinese Physics B, {\bf 23}, 110302 (2013)

\bibitem{endoEtAl2014}
T. Endo, N. Konno, E. Segawa, M. Takei: A one-dimensional Hadamard walk with one defect,
Yokohama Mathematical Journal, {\bf 60}, 49-90 (2014)

\bibitem{endokonno2014}
T. Endo and N. Konno: The stationary measure of a space-inhomogeneous quantum walk on the line, Yokohama Mathematical Journal, {\bf 60}, 33-47 (2014)

\bibitem{endoEtAl2015}
S. Endo, T. Endo, N. Konno, E. Segawa, and M. Takei:
Limit theorems of a two-phase quantum walk with one defect, Quantum Information and Computation, {\bf 15}, 1373-1396 (2015)

\bibitem{endoEtAl2016}
S. Endo, T. Endo, N. Konno, E. Segawa, and M. Takei:
Weak limit theorem of a two-phase quantum walk with one defect,
 Interdisciplinary Information Sciences, {\bf 22}, 17-29 (2016)
 
%\bibitem{endo}
%T. Endo and N. Konno:
%The time-averaged limit measure of the Wojcik model, Quantum Information and Computation, {\bf 15}, 0105-0133 (2015)

%\bibitem{endot}
%T. Endo and N. Konno:
%Weak Convergence of the Wojcik model, Yokohama Mathematical Journal, {\bf 61}, 87-111 (2015)

\bibitem{endokawaikonno2016}
T. Endo, H. Kawai, and N. Konno:
Stationary measures for the three-state Grover walk with one defect in one dimension, Suurikaisekikokyuroku, {\bf 2010}, 45-55 (2016)

\bibitem{endobusekonno2015}
T. Endo, N. Konno, ans H. Obuse:
Relation between two-phase quantum walks and the topological invariant, arXiv 1511.04230, 1-50 (2015)

%\bibitem{flajolet}
%P. Flajolet and R. Sedgewick: Analytic combinatorics, Cambridge University Press (2009)

%\bibitem{fu}
%L. Fu and C. L. Kane:
%Superconducting Proximity Effect and Majorana fermions at the surface of a topological insulator, Physical %Review Letters, {\bf 100}, 096407 (2008)

\bibitem{schudoEtAl2004}
G. Grimmett, S. Janson, and P. F. Scudo:
Weak limits for quantum random walks, Physical Review E, {\bf 69}, 026119 (2004)

\bibitem{inuiEtAl2003}
N. Inui, Y. Konishi, and N. Konno:
Localization of two-dimensional quantum walks, Physical Review A, {\bf 69}, 052323 (2003)

\bibitem{joyemerkli2010}
A. Joye and M. Merkli:
Dynamical localization of quantum walks in random environments, Journal of Statistical Physics, {\bf 140},1025-1053 (2010)

\bibitem{kawaiEtAl2017}
H. Kawai, T. Komatsu, and N. Konno:
Stationary measures of three-state quantum walks on the one-dimensional lattice, Yokohama Mathematical Journal, {\bf 63}, 59-74 (2017)

\bibitem{komatsukonno2017}T. Komatsu and N. Konno: Stationary amplitudes of quantum walks on the higher-dimensional integer lattice, Quantum Information Processing, {\bf 16}, 291 (2017)

\bibitem{kitagawaEtAl2010}
T. Kitagawa, M. S. Rudner, E. Berg, and E. Demler:
Exploring topological phases with quantum walks, Physical Review A, {\bf 82}, 033429 (2010)

%\bibitem{koyoo2013}
%C. K. Ko and H. J. Yoo:
%The generator and quantum Markov semigroup for quantum walks, Kodai Mathematical Journal, {\bf 36},363-385 (2013)

\bibitem{konnoweak2005}
N. Konno:
A new type of limit theorems for the one-dimensional quantum random walk, Journal of the Mathematical Society of Japan, {\bf 57}, 1179-1195 (2005)

%\bibitem{konnopath2005}
%N. Konno:
%A path integral approach for disordered quantum walks in one dimension, Fluctuation and Noise Letters, {\bf 5}, 529-537 (2005)

\bibitem{konno2008} 
N. Konno:
Quantum walks, Quantum Potential Theory, Lecture Notes in Mathematics, {\bf 1954}, 309-452 (2008)
 
\bibitem{konno2009}
N. Konno:
One-dimensional discrete-time quantum walks on random environments, Quantum Information Processing, {\bf 8}, 387-399 (2009)

\bibitem{konno2010}
N. Konno:
Localization of an inhomogeneous discrete-time quantum walk on the line, Quantum Information Processing, {\bf 9}, 405-418 (2010)

\bibitem{konnoluczaksegawa2013}
N. Konno, T. Luczak, and E. Segawa:
Limit measures of inhomogeneous discrete-time quantum walks in one dimension, Quantum Information Processing, {\bf 12}, 33-53 (2013)


\bibitem{konnoyoo2013}
N. Konno and H. J. Yoo:
Limit theorems for open quantum random walks, Journal of Statistical Physics, {\bf 150}, 299-319 (2013)

\bibitem{konno2014}
N. Konno:
The uniform measure for discrete-time quantum walks in one dimension, Quantum Information Processing,
{\bf 13}, 1103-1125 (2014)

\bibitem{konnotakei2015}
N. Konno and M. Takei:
The non-uniform stationary measure for discrete-time quantum walks in one dimension, Quantum Information and Computation, {\bf 15}, 1060-1075 (2015)

\bibitem{morioka2019}
H. Morioka:
Generalized eigenfunctions and scattering matrices for position-dependent quantum walks, 
Reviews in Mathematical Physics, 1-37  (2019) 

\bibitem{sadowskiEtAl2016}
P. Sadowski, J. Adam Miszczak, and M. Ostaszewski:
Lively quantum walks on cycles, 
Journal of Physics A: Mathematical and Theoretical, {\bf 49} 375302 (2016)

\bibitem{kempeEtAl2003}
N. Shenvi, J. Kempe, and K. B. Whaley:
A quantum random walk search algorithm, Physical Review A, {\bf 67}, 052307 (2003)

\bibitem{wan2015}
C. Wang, X. Lu, W. Wang:
The stationary measure of a space-inhomogeneous three-state quantum walk on
the line, Quantum Information Processing, {\bf 14}, 867-880 (2015)

\bibitem{wojcik2012}
A. Wojcik, T. Luczak, P. Kurzynski, A. Grudka, T. Gdala, and M. Bednarska-Bzdega:
Trapping a particle of a quantum walk on the line, Physical Review A, {\bf 85}, 012329 (2012)

\end{thebibliography}
\end{document}